\documentclass[a4paper, twoside, fontsize=9pt, twocolumn]{scrartcl}   
\pdfoutput=1

\usepackage[a4paper, top=88pt, bottom=88pt, left=50pt, right=50pt, headsep=16pt, footskip=28pt]{geometry}  

\usepackage{amsfonts,amssymb,amsmath,amsthm,amstext,amssymb,amsopn,mathtools,nicefrac,xfrac}
\usepackage[authoryear,sort,round]{natbib}  
\usepackage[utf8]{inputenc}                 
\usepackage{booktabs}                       
\usepackage{nicefrac}                       
\usepackage{microtype}                      
\usepackage{graphics,graphicx}              
\usepackage{subcaption}                     
\usepackage{lipsum}                         
\usepackage{tabularx}

\usepackage[boxruled,linesnumbered]{algorithm2e} 
\SetKwComment{Comment}{\%}{}                     
\SetKwInput{KwInput}{Input}
\SetKwInput{KwOutput}{Output}

\usepackage{verbatim}                       
\usepackage{appendix}                       
\usepackage{bm}                             
\usepackage{array}                          
\newcolumntype{C}[1]{>{\centering\arraybackslash}m{#1}}  
\usepackage[usenames,dvipsnames]{xcolor}    
\usepackage{hyperref}                       
\hypersetup{colorlinks=true,linkcolor=Maroon,filecolor=Magenta,urlcolor=Blue,citecolor=RoyalBlue}       
\usepackage[noabbrev,capitalise,nosort,nameinlink]{cleveref}  
\usepackage{bbm}                            
\usepackage{mathtools}                      
\usepackage{soul} 

\newcommand*{\arXiv}[1]{\bgroup\color{blue}\href{https://arxiv.org/abs/#1}{arXiv:#1}\egroup}
\newcommand*{\doi}[1]{\bgroup\color{blue}\href{https://doi.org/#1}{doi:#1}\egroup}
\newcommand*{\email}[1]{\bgroup\color{blue}\href{mailto:#1}{#1}\egroup}
\renewcommand*{\url}[1]{\bgroup\color{blue}\href{#1}{#1}\egroup}
\usepackage{enumitem, moreenum}
\setlist[enumerate]{nosep}
\setlist[itemize]{nosep}
\usepackage{mleftright} \mleftright
\renewcommand{\qedsymbol}{$\blacksquare$}
\renewenvironment{proof}[1][\proofname]{\noindent{\bfseries\sffamily #1.} }{\hfill\qedsymbol\medskip}
\usepackage[textfont={small}, labelfont={sf,bf,small},format=plain,indention=0cm]{caption}
\DeclareCaptionLabelSeparator{figlabelsep}{\,\,\,}
\usepackage{scrlayer-scrpage, xhfill}
\automark[section]{section}
\setkomafont{pageheadfoot}{\normalcolor\sffamily}
\setkomafont{pagenumber}{\normalfont\normalsize\sffamily}
\clearpairofpagestyles
\let\oldtitle\title
\renewcommand{\title}[1]{\oldtitle{#1}\newcommand{\theshorttitle}{#1}}
\newcommand{\shorttitle}[1]{\renewcommand{\theshorttitle}{#1}}
\let\oldauthor\author
\renewcommand{\author}[1]{\oldauthor{#1}\newcommand{\theshortauthor}{#1}}
\newcommand{\shortauthor}[1]{\renewcommand{\theshortauthor}{#1}}
\cohead{\xrfill[0.525ex]{0.6pt}~\theshorttitle~\xrfill[0.525ex]{0.6pt}}
\cehead{\xrfill[0.525ex]{0.6pt}~\theshortauthor~\xrfill[0.525ex]{0.6pt}}
\newcommand{\theabstract}[1]{\par\bgroup\noindent\textbf{\textsf{Abstract.}} #1\egroup}
\newcommand{\thekeywords}[1]{\par\smallskip\bgroup\noindent\textbf{\textsf{Keywords.}}\newcommand{\and}{ $\bullet$ } #1\egroup}
\newcommand{\themsc}[1]{\par\smallskip\bgroup\noindent\textbf{\textsf{2020 Mathematics Subject Classification.}}\newcommand{\and}{ $\bullet$ } #1\egroup}
\newcommand*{\affilref}[1]{\ref{affiliation#1}}
\newcommand*{\affiliation}[3]{
	\footnotetext[#1]{\label{affiliation#2} #3}
}
\usepackage{siunitx} 
\usepackage{tikz}
\usetikzlibrary{arrows.meta,positioning}

\numberwithin{equation}{section}
\numberwithin{figure}{section}
\numberwithin{table}{section}

\theoremstyle{definition}

\crefname{assumption}{Assumption}{Assumptions}
\Crefname{assumption}{Assumption}{Assumptions}

\newcommand*{\Rin}{R_{\text{in}}}
\newcommand*{\Rout}{R_{\text{out}}}
\newcommand*{\Rx}{R_{\text{X}}}
\newcommand*{\Zx}{Z_{\text{X}}}
\newcommand*{\Rnose}{R_{\text{nose}}}
\newcommand*{\dgap}{S_{\text{gap}}}
\newcommand*{\Rs}{R_{\text{s}}}

\definecolor{darkspringgreen}{rgb}{0.09, 0.45, 0.27}
\definecolor{amber(sae/ece)}{rgb}{1.0, 0.49, 0.0}

\title{Real-time virtual circuits for plasma shape control via neural network emulators: experimental demonstration on MAST Upgrade}
\shorttitle{Deployment of real-time virtual circuits on MAST-U}
\author{
    N. C. Amorisco\textsuperscript{\affilref{UKAEA}}
    \and
    K. Pentland\textsuperscript{\affilref{UKAEA}}
    \and
    A. Agnello\textsuperscript{\affilref{Hartree}}
    \and
    G. K. Holt\textsuperscript{\affilref{UKAEA}}
    \and
    A. Ross\textsuperscript{\affilref{Hartree}}
    \and
    M. J. Marshall\textsuperscript{\affilref{Hartree}}
    \and
    E. Jones\textsuperscript{\affilref{UKAEA}}
    \and
    G. McArdle\textsuperscript{\affilref{UKAEA}}
    \and
    C. Vincent\textsuperscript{\affilref{UKAEA}}
    \and
    T. Nunn\textsuperscript{\affilref{UKAEA}}
    \and
    M. Kochan\textsuperscript{\affilref{UKAEA}}
    \and
    P. Cavestany\textsuperscript{\affilref{Hartree}}
    \and
    A. Garrod\textsuperscript{\affilref{Hartree}}
    \and
    S. Pamela\textsuperscript{\affilref{UKAEA}}
    \and
    J. Buchanan\textsuperscript{\affilref{UKAEA}}
    \and 
    the MAST Upgrade Team\textsuperscript{\affilref{MASTU}}
}
\shortauthor{N. C. Amorisco et al.}
\date{\today\\[0.3em]}

\begin{document}
\maketitle

\affiliation{1}{UKAEA}{United Kingdom Atomic Energy Authority, Culham Campus, Abingdon, Oxfordshire, OX14 3DB, United Kingdom\newline (\email{nicola.amorisco@ukaea.uk})}
\affiliation{2}{Hartree}{STFC Hartree Centre, Sci-Tech Daresbury, Keckwick Lane, Daresbury, Warrington, WA4 4AD, United Kingdom}
\affiliation{3}{MASTU}{See author list of J.R. Harrison at al 2026 Nucl. Fusion 66 116005}


\begin{abstract}\small
    \theabstract{
    Conventional plasma shape control in tokamaks relies on virtual circuits (VCs) that are computed offline from linearisations around a small, tailored number of reference equilibria, and deployed as expertly prepared schedules during the discharge. Here, we report on the first experimental deployment of real-time VCs. We replace pre-set look up tables with VCs updated in real time using surrogates of the plasma response. Both the existing control architecture and the interpretability of VC-based control are retained.
    
    Previous work showed that neural network emulators can produce accurate VCs, and validated their performance in closed-loop shape control simulations. Here, we report their first experimental validation on MAST Upgrade (MAST-U). Dedicated experiments spanning different scenarios, including prescribed shape perturbations, feedback-driven divertor-leg motion, and strongly evolving plasma configurations, show that real-time VCs can realise plasma shape-control tasks within the MAST-U plasma control system.

    These results establish the experimental feasibility of real-time linearisations as a practical extension of conventional plasma shape control in tokamaks. The present implementation demonstrates a central step towards a simpler control workflow, in which manually constructed, phased VC schedules are replaced by VCs generated automatically online from a trained surrogate model, without scenario-specific retraining.
    }
    \thekeywords{%
        {Virtual circuits}%
        \and
        {Real-time control}%
        \and%
        {Neural-network emulators}%
        \and%
        {Plasma control systems}%
        \and%
        {MAST Upgrade}%
    }
   
\end{abstract}

\section{Introduction} \label{sec:intro}

Accurate control of the plasma position and shape is fundamental to the operation of magnetic confinement fusion experiments, enabling the exploration of high-performance operating scenarios while maintaining safe plasma operation \citep[e.g.,][]{ariola2016}. 
As fusion devices move towards higher power operation and increasingly demanding engineering constraints, there is growing interest in the application of artificial intelligence (AI) and machine learning to plasma control, with the aim of improving controller performance, reducing the burden of manual controller design and tuning, and enabling increasingly autonomous operation.

Applications of AI to tokamak plasma control have significantly progressed in the 2020s, with approaches spanning several levels of the control architecture. Neural networks (NNs) have provided real-time estimates for vertical stability control on DIII-D and for fast and adaptive vertical position control on EAST \citep{sammuli2021,song2024,rui2025}, building on earlier neural-network-assisted equilibrium control experiments \citep{bishop1995,windsor1997}. Learned predictors and policies have also been deployed on DIII-D for plasma performance control, tearing instability avoidance and prevention of H--L back-transitions \citep{char2023,seo2024,orozco2022}.
Reinforcement learning has been used to develop policies that map measurements and control targets either to high-level control requests or directly to actuator commands \citep{de_tommasi2022,degrave2022,seo2024,tracey2024,benlarbi2024,char2023,wang2025} and used to control plasmas on TCV, HL-3, and DIII-D \citep{degrave2022,tracey2024,wu2025,subbotin2026}.

\begin{figure*}[t!]
    \begin{subfigure}{0.99\linewidth}
        \centering
        \includegraphics[width=0.99\textwidth]{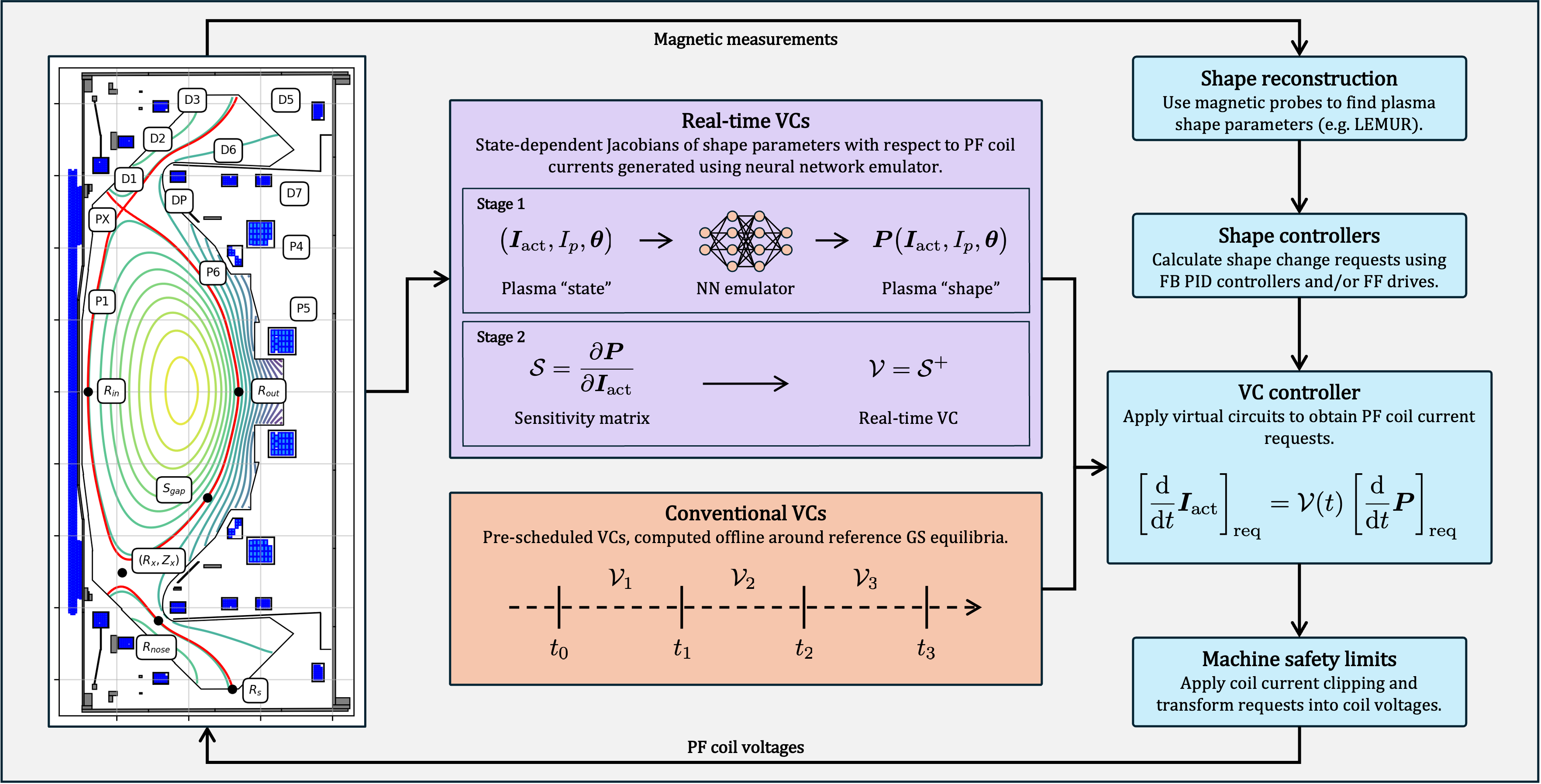}
    \end{subfigure}
    \caption{Schematic comparison between conventional virtual-circuit operation and RTVC operation. Conventional operation uses precomputed linearisations scheduled by discharge phase, whereas RTVCs generate the local linearisation from the instantaneous plasma state. The shape reconstruction, shape/VC controllers, and machine safety limits within the PCS interface are unchanged.}
    \label{fig:concept}
\end{figure*}

While these developments demonstrate the potential of AI to enhance
plasma control and operation, their routine use within a plasma control system (PCS) must
satisfy considerations beyond control robustness. In particular,
interpretability and operational familiarity remain important, while
retraining learned controllers around new scenarios can be costly.
In this work, we demonstrate a different approach designed to address
these concerns. Rather than replacing the feedback controller, we use
neural network emulators to update the local plasma response
linearisations from which virtual circuits (VCs) are derived. Shape errors and 
actuator requests therefore remain connected through the established
VC decomposition, retaining the structure and operational familiarity
of conventional shape control. Because the emulators are trained over
a broad operating space rather than for individual discharge
scenarios, the same models can provide real-time VCs across
different plasma conditions without scenario-specific retraining or
manually prepared VC schedules. Compared with static, pre-set schedules, these real-time updates keep
the VCs matched to the evolving plasma response, maintaining the
intended orthogonality between shape-control parameters and allowing the
existing controller to adapt its actuator response when the plasma
departs unexpectedly from the planned trajectory.

The timeliness of this approach is reinforced by a contemporaneous experimental demonstration of an inverse Grad-Shafranov neural network for tokamak magnetic control \citep{wang26}. That method maps requested plasma shapes and real-time estimates of the plasma state directly to PF-coil current references, which are then tracked by lower-level current controllers. RTVC addresses the same broad challenge from a complementary direction: rather than learning the inverse mapping from shape requests to coil currents, it learns the forward plasma response, differentiates this model to construct local VCs, and retains explicit feedback on the reconstructed plasma-shape error. The emergence of these complementary approaches highlights real-time, learned equilibrium models as an increasingly active direction in tokamak shape control.

The methodological and validation foundations of RTVC were established in a sequence of earlier studies. \citet{agnello2024} introduced a framework for efficiently
generating synthetic Grad-Shafranov (GS) equilibria using the FreeGSNKE
free-boundary equilibrium solver \citep{amorisco2024}.
\citet{ross2026} applied this framework at scale to produce a large
equilibrium library spanning a broader range of configurations than
those explored in previous campaigns on MAST Upgrade
\citep[MAST-U;][]{Fishpool2013,MCARDLE2020111764,ryan2023}.
The resulting library was used to train plasma-shape emulators, whose
Jacobians were used to construct VCs, which were then tested against physics-based
reference VCs in static validation.
Subsequent closed-loop simulations using the FreeGSNKE Pulse Design
Tool (FPDT) \citep{pentland2026a} showed that these real-time
VCs delivered the requested shape corrections across a broad region
of the MAST-U operating space \citep{pentland2026b}. Most recently,
the emulators were incorporated into a real-time
virtual circuit (RTVC) server and control
algorithm and integrated into the MAST-U PCS. Playback, performance,
and non-actuating ``piggyback'' tests verified real-time execution
and consistency with the offline emulator implementation
\citep{marshall2026}.
    
Here, we present the first experimental demonstration of RTVCs on MAST-U. 
The experiments span progressively more demanding plasma scenarios,
    from control of conventional plasma shapes, through prescribed shape
    perturbations and feedback-driven divertor leg motion, to strongly
    evolving plasma configurations. Together, they establish the
    feasibility of integrating RTVCs within an existing control
    architecture and applying the same trained models across distinct
    scenarios without scenario-specific retraining or manually prepared
    VC schedules. They also provide a first experimental assessment of
    the resulting shape control performance and its present limitations.

\section{Experimental methodology} \label{sec:RTVC}

\subsection{Real-time VCs}

    VCs are conventionally constructed by linearising the controlled
    plasma shape parameters with respect to the poloidal field (PF) coil currents around a
    reference GS equilibrium, and inverting, typically through a
    pseudoinverse, the resulting Jacobian matrix. Each VC is therefore a vector of PF
    coil current perturbations that, to first order, change one local
    shape parameter, such as the X-point position, by a prescribed amount
    while suppressing changes in the other controlled parameters \citep[see, e.g.,][]{wai26}.

    In conventional MAST-U operation, these linearisations are computed
    offline at a limited set of reference equilibria sampled along the
    planned plasma trajectory. The corresponding VCs are then assigned
    manually to predefined phases of the pulse. Selecting the reference
    equilibria and the transition times between their VCs is therefore an
    integral part of expert scenario preparation. Within each phase, the
    scheduled VCs remain fixed. As the plasma evolves, or departs
    unexpectedly from the planned trajectory, its local shape response
    can differ from that of the reference equilibrium. The scheduled VCs
    then become increasingly mismatched, potentially degrading the intended
    decoupling between shape control channels and reducing control
    effectiveness.

In the low-latency approach introduced by \citet{ross2026}, neural-network emulators predict the plasma shape from a reduced description of its instantaneous state. The reduced plasma state comprises the currents in the twelve active poloidal-field (PF) coils of MAST-U, the total plasma current, and a small number of parameters describing the plasma current-density profile. From these inputs, the emulators predict seven shape parameters routinely used in MAST-U control: the inner and outer midplane radii $(\Rin,\Rout)$; the lower X-point position $(\Rx,\Zx)$; the ``squareness gap'' $\dgap$, defined between the plasma core and a specified point on the main-chamber wall; the lower outboard strike-point position $\Rs$; and $\Rnose$, the radial position at which the lower divertor leg intersects a virtual line across the divertor entrance. Local shape-response matrices are obtained by differentiating the emulator predictions with respect to the active PF coil currents, using either finite differences or automatic differentiation, and the corresponding VCs are then constructed from these matrices. Further details and definitions are provided by \citet{ross2026}.

While the plasma and PF coil currents are provided in real time by the plasma control system (PCS), in this experimental deployment the profile parameters are prescribed. We use reconstructed profile parameters from prior representative discharges. Also, the implementation of RTVC presented here does not account for the currents induced in the passive structures within the emulator state.

\subsection{Integration into the MAST-U plasma control system}

    The NN emulators are deployed through a dedicated real-time inference
    server integrated with the MAST-U PCS; implementation details are
    given in \citet{marshall2026}. At each VC update, the server receives the
    plasma current, PF coil currents, and plasma profile parameters,
    evaluates the emulators, and returns the local shape response
    Jacobian. The PCS computes its pseudoinverse and uses the result to
    replace selected entries of the conventional VC matrix. The
    downstream controller, gains, and machine protection handling remain
    unchanged, allowing conventional and real-time VCs to be selected by
    shape parameter and pulse phase, as illustrated in
    \cref{fig:concept}.

    Because the real-time inference engine does not yet support automatic
    differentiation, the Jacobian is evaluated by finite differences of
    the emulator outputs. \citet{ross2026} showed that this produces VCs
    of comparable accuracy to those obtained using automatic
    differentiation.

    Before active plasma control, the complete deployed workflow was
    validated through software-level tests, offline discharge playback,
    and comparison with the TensorFlow emulator implementation in the
    FPDT. It was then operated in non-actuating ``piggyback'' mode during
 MAST-U discharges, during which the conventional controller continued to
    actuate the plasma, while RTVC recorded the PF coil requests it would
    have issued from the same measurements \citep{marshall2026}. 
    This validation allows the experiments in
    \cref{sec:experiments} to focus on the closed-loop plasma response to
    the real-time VCs.

    The MAST-U shape controller runs at \SI{10}{\kilo\hertz} using
    the latest available VC, with RTVC supplying an updated matrix every
    \SIrange{5}{5.6}{\milli\second}.

\section{Experimental demonstration} \label{sec:experiments}

We designed experiments to validate complementary aspects of the proposed RTVC framework, progressing from compatibility with plasma scenarios routinely run on MAST-U to more dynamic plasma shape evolution for which RTVCs are principally intended. 
These experiments are not intended to demonstrate superior control performance relative to conventional VCs. Such a comparison would require a dedicated campaign of discharges specifically aimed at a systematic comparison, which was not possible within the available machine time. Instead, the purpose of these experiments has been to assess whether RTVCs could reliably realise plasma shape control tasks of increasing complexity while preserving the existing controller architecture.

\begin{figure*}[t!]
    \centering
    \begin{subfigure}{0.99\linewidth}
        \includegraphics[width=0.99\textwidth]{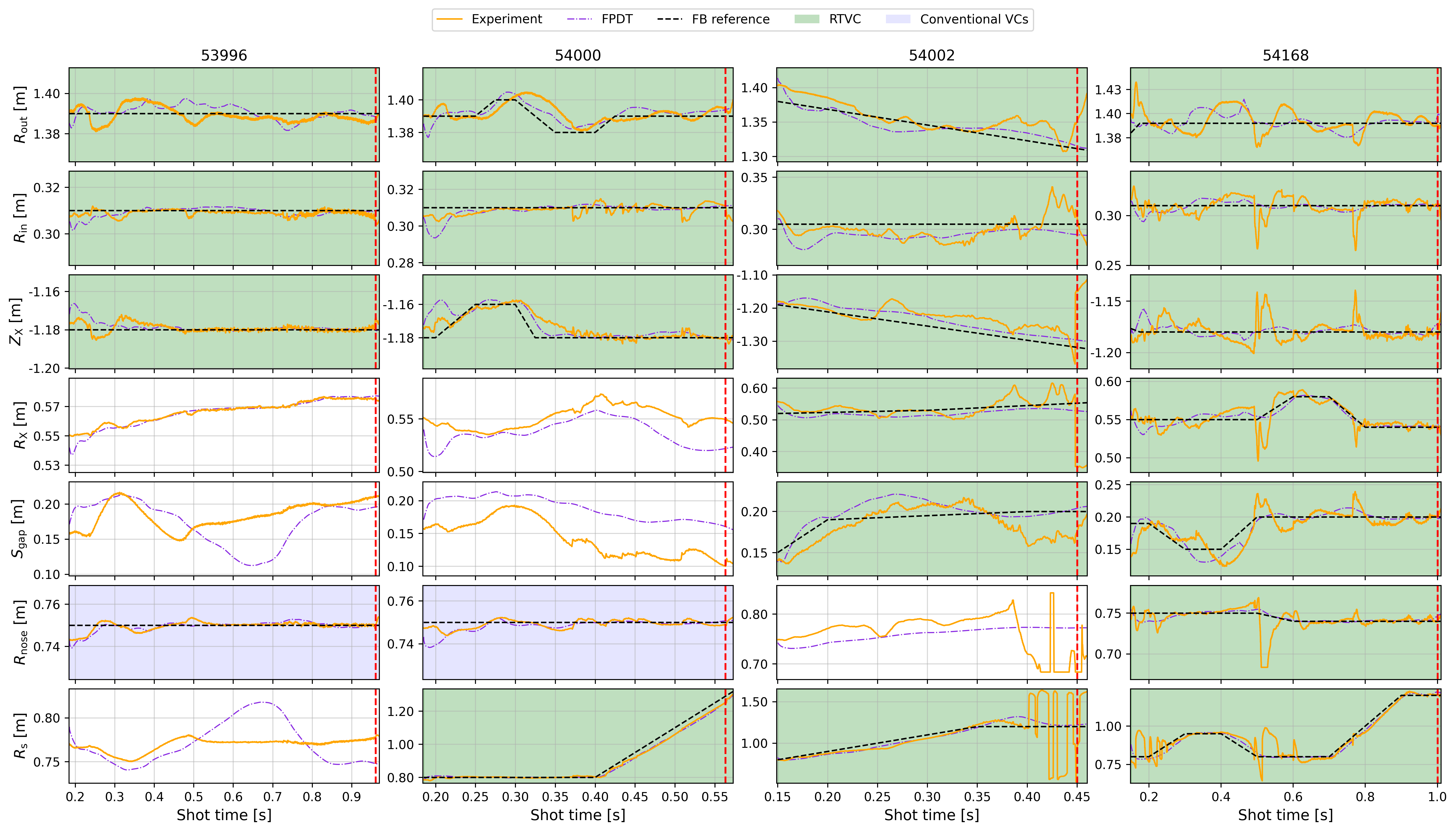}
    \end{subfigure}
    \caption{Evolution of the shape parameters in each RTVC-controlled MAST-U shot. Shown are the experimentally observed values recorded in real-time by the LEMUR algorithm (orange), the FPDT simulations made prior to the discharge (dashed-dot purple), and the pre-programmed feedback reference waveforms (dashed black). The shading indicates periods where feedback control was switched on (green with RTVC, lilac with Conventional VCs). Vertical dashed red lines indicate when the plasma went into ramp-down.
    }
    \label{fig:experimental_summary}
\end{figure*}

Table~\ref{tab:shots_overview} and Figure~\ref{fig:experimental_summary} provide an overview of the RTVC-controlled discharges. A green background in Figure~\ref{fig:experimental_summary} indicates that the shape parameter is feedback controlled by RTVC: the corresponding feedback reference is displayed with a black dashed line, while the measured shape evolution as reconstructed in real time by the LEMUR algorithm \citep{kochan2023} is shown in orange. Shape parameters controlled using RTVC-provided VCs are also listed in the middle column of Table~\ref{tab:shots_overview}. 
Together with the real time shape measurements, Fig.~\ref{fig:experimental_summary} also displays the pre-shot simulation results obtained for the same discharges using the emulators within the FPDT.

\begin{table*}[t]
\centering
\small
\caption{Overview of the experimental validation objectives, RTVC controlled shape parameters, and shape parameters used in the matrix inversion.}
\label{tab:shots_overview}
\begin{tabularx}{\textwidth}{|c|X|X|X|}
\hline
\textbf{Shot} & \textbf{Objective} & \textbf{RTVC controlled shape pars.} & \textbf{Included in inverse matrix} \\
\hline\hline
53996 & Constant shape control &
($\Rout$, $\Rin$, $\Zx$) &
($\Rout$, $\Rin$, $\Zx$, $\dgap$, $\Rnose$) \\
54000 & Minor core shape perturbations and sweep divertor-leg &
($\Rout$, $\Rin$, $\Zx$, $\Rs$) &
($\Rout$, $\Rin$, $\Zx$, $\dgap$, $\Rnose$, $\Rs$) \\
54002 & Increase elongation and sweep divertor-leg &
($\Rout$, $\Rin$, $\Zx$, $\Rx$, $\dgap$, $\Rs$) &
($\Rout$, $\Rin$, $\Zx$, $\Rx$, $\dgap$, $\Rs$) \\
54168 & Simultaneous control of all seven shape parameters and shape perturbations &
($\Rout$, $\Rin$, $\Zx$, $\Rx$, $\dgap$, $\Rnose$, $\Rs$) &
($\Rout$, $\Rin$, $\Zx$, $\Rx$, $\dgap$, $\Rnose$, $\Rs$) \\
\hline
\end{tabularx}
\end{table*}

\subsection{Routine plasma control}

Shot $53996$ was designed to conservatively test compatibility of the RTVC framework with routine MAST-U shape control, and was based on reference discharge 53320 -- a \SI{750}{\kilo\ampere} double-null, conventional-divertor scenario, heated with two neutral beams and with no plasma shape evolution.
In this shot, $\Rout$, $\Rin$, $\Zx$ and $\Rnose$ are actively controlled. RTVC takes control of the core shape ($\Rout$, $\Rin$, $\Zx$) at t=$185~\mathrm{ms},$ while $\Rnose$ control is performed using a conventional VC schedule (as highlighted by the lilac shading), the same used in the reference shot. The real-time VCs derived by RTVC are obtained as pseudoinverse of the Jacobian including all of $\Rout$, $\Rin$, $\Zx$, $\dgap$, $\Rnose$.
Although $\dgap$ is not actively controlled in this shot, including it in the pseudoinversion discourages the RTVC control actions from producing unintended changes in this parameter.

The discharge reached its programmed duration and the RTVC-controlled shape parameters remained close to the requested waveforms throughout the controlled phase, demonstrating the use of both real-time and conventional VCs together.

\subsection{Prescribed shape perturbations and divertor-leg motion}

Shot $54000$ tested whether RTVC could enact deliberate shape changes. 
The scenario was constructed from the successful $53996$ configuration, but introduced prescribed trapezoidal perturbations to the core shape through $\Rout$ and $\Zx$, and a partly simultaneous feedback-driven divertor-leg sweep, aiming for the Super-X configuration \citep{valanju2009, morris2014,morris2018}.
The $\Rs$ reference was designed to sweep the leg, using RTVC, from approximately \SI{0.8}{\metre} to \SI{1.4}{\metre} between \SI{400}{\milli\second} and \SI{600}{\milli\second}. 
As in shot $53996$, $\Rnose$ was retained under the same pre-set VC schedule as the original reference shot, while $\dgap$ was included in the inverse but not actively controlled by the RTVC.

The measured evolution, shown in \cref{fig:experimental_summary}, demonstrates successful tracking of the imposed trapezoidal shape perturbations in both $\Rout$ and $\Zx$, and accurate enacted motion of the divertor leg. 
In particular, the $\Rs$ evolution follows the requested outward sweep until the discharge terminates at \SI{563}{\milli\second} following a Langmuir probe protection trip. 
The termination was not attributed to RTVC operation, but was consistent with an operational limitation encountered during Super-X attempts in the same experimental session, thought to be associated with a possible bias in the real-time LEMUR estimate of $\Rs$ near the outer divertor tiles. Prior to the termination, RTVC successfully enacted the requested changes in both the core shape and divertor-leg geometry.

\subsection{Dynamic plasma shape evolution}

Shot $54002$ was purposely designed to be more demanding, targeting a dynamic transition towards a plasma shape that has remained difficult to maintain reliably and control robustly on MAST-U: an elongated plasma with a Super-X leg configuration. This is the dynamic scenario that the RTVC paradigm is best posed to address: a highly evolving plasma shape in which the local linearisations change significantly during the discharge, making control design via conventional pre-set VC phases especially challenging. 

In this shot, all shape control was performed using RTVC, which controlled six out of seven shape parameters. In order to avoid the Langmuir probe protection limit encountered in shot $54000$ the $\Rs$ sweep was limited to \SI{1.2}{\metre}. 
RTVC control was enabled earlier than in the previous shots, at \SI{150}{\milli\second}.
The core shape parameters were driven towards a higher plasma elongation configuration.

\begin{figure}[t!]
    \begin{subfigure}{0.99\linewidth}
        \centering
        \includegraphics[width=0.99\textwidth]{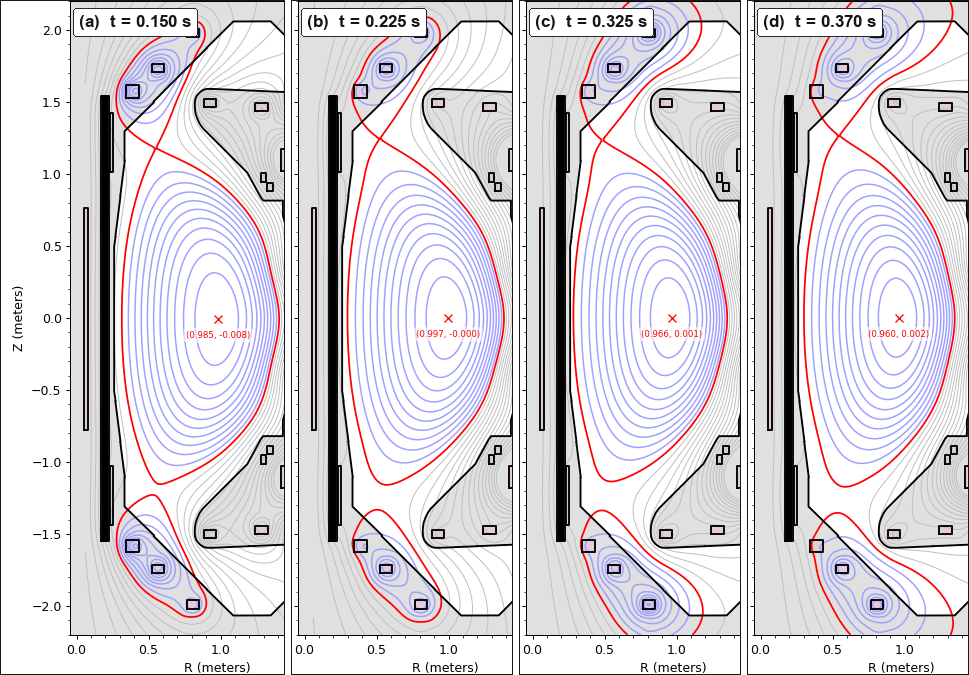}
    \end{subfigure}
    \caption{Poloidal magnetic flux evolution in shot $54002$ during the RTVC-controlled phase, generated using EFIT\texttt{++}. Snapshots are shown at RTVC activation, \SI{140}{\milli\second} after activation, and \SI{220}{\milli\second} after activation, illustrating the evolution from a conventional configuration towards the higher elongation with near-super-X configuration.
    }
    \label{fig:shot54002}
\end{figure}

\Cref{fig:experimental_summary} shows RTVC successfully enacting the controlled ramp towards higher elongation through changes in both $\Rout$ and $\Zx$, together with the prescribed adjustment to $\dgap$. The divertor-leg sweep is also completed successfully, after which the strike-point position $\Rs$ is maintained without triggering the real-time protection system. At approximately \SI{380}{\milli\second}, interaction with the nose results in a loss of shape control. This interaction was not reproduced in the pre-shot FPDT simulation. Although $\Rx$ was controlled in this shot, unlike in the reference shot $\Rnose$ was not directly constrained. A discrepancy between the modelled and realised plasma responses could therefore produce an uncorrected displacement of the plasma nose while the actively controlled parameters remained close to their targets. Nevertheless, the requested shape transition had already been completed during the preceding RTVC-controlled interval.
This is independently corroborated by the offline EFIT++ reconstructions of the same shot, shown in \cref{fig:shot54002}, reproducing the substantial evolution in triangularity, elongation, and leg position inferred in real time by LEMUR and driven by RTVC control.

\subsection{Simultaneous control of seven shape parameters}

Shot $54168$ was an exploratory stress test of simultaneous
feedback control of all seven shape parameters available to
RTVC. Such operation is not routinely attempted on MAST-U
because the shape responses are strongly cross-coupled and
the resulting pseudoinversion can be poorly conditioned. Unlike
the preceding experiments, all seven parameters were included
in the RTVC inverse and assigned active feedback control.

RTVC was enabled at \SI{150}{\milli\second}, and the discharge
reached its programmed ramp-down at \SI{1}{\second}. As shown
in \cref{fig:experimental_summary}, transient departures from
the requested shape occurred around \SI{0.50}{\second} and
\SI{0.77}{\second}. The shape parameters subsequently returned
towards their references, but tracking was markedly less smooth
than in the preceding experiments. This result motivates the
analysis of the VC inversion presented in
\cref{sec:discussion}.


%


\section{Discussion and outlook}
\label{sec:discussion}

This work demonstrates that real-time, emulator-derived local plasma
linearisations can be deployed in closed-loop control of MAST-U plasmas.
Across discharges involving a constant shape, prescribed perturbations,
feedback-driven divertor leg motion, and a strongly evolving configuration, 
the same RTVC implementation and emulators were used without
scenario-specific retraining and without replacing the established MAST-U
shape control architecture. Shot $54168$ extended this test to the simultaneous
control of all seven available shape parameters. The campaign therefore
establishes the experimental feasibility of RTVC across distinct control
tasks.

The experiments also clarify the consequences of the approximations made in
the present emulator configuration. Currents induced in the passive conducting
structures are not included, and the plasma current density profile is
represented by prescribed parameters rather than real-time estimates. The
successful discharges show that these approximations did not preclude shape
control in the states tested, consistent with the preceding simulations
\citep{pentland2026b}. They do not imply that passive currents or profile
evolution are negligible in general, particularly during rapidly evolving
phases. Including estimates of the passive current state and real-time plasma
profiles would provide a more complete description of the instantaneous
equilibrium and is a priority for future development.

\begin{figure}
    \centering
    \includegraphics[width=.9\columnwidth]{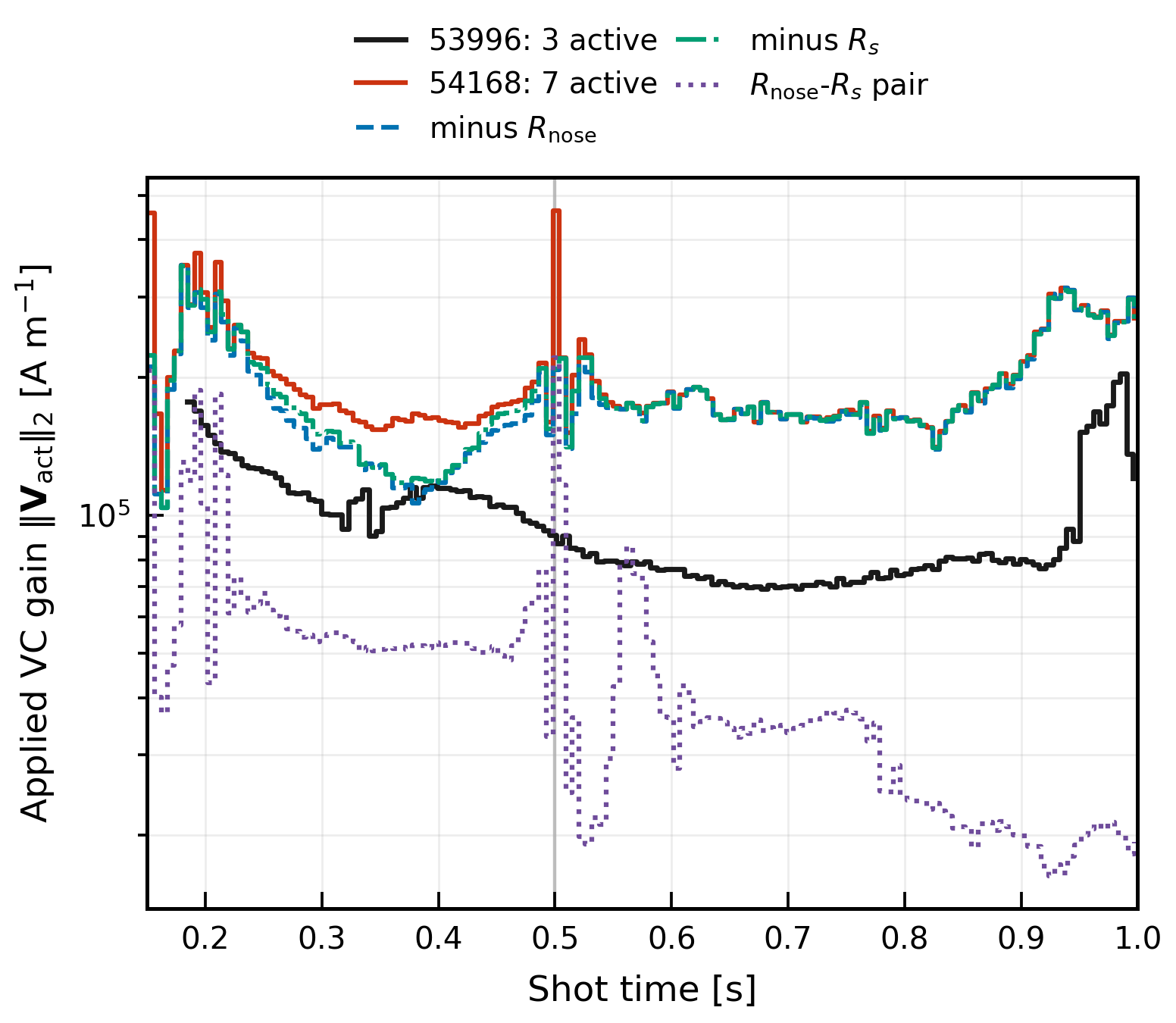}
    \caption{
    The VC norm $\|\mathcal{V}_{\text{act}}\|_2$ of the RTVC-controlled shape parameters in shots $53996$ (solid black) and $54168$ (solid red). 
    Recall the shape matrix in shot $53996$ included $\Rout$, $\Rin$, $\Zx$, $\dgap$ and $\Rnose$, while RTVC feedback was applied to the first three ($\Rnose$ being controlled with a static VC)  
    The pseudoinversion with all seven shape parameters in shot $54168$ is compared with offline diagnostic inversions omitting either $\Rnose$ (dashed blue) or $\Rs$ (dashed green), and with the $\Rnose$--$\Rs$ pair alone (dotted purple). 
    The diagnostic inversions are shown for comparison and were not used to control the discharge. 
    A vertical dotted line is denoted to mark the local amplification event at approximately \SI{0.5}{\second}.
    }
    \label{fig:inverse_amplification}
\end{figure}

The present RTVC implementation also retained the Moore--Penrose pseudoinverse $\mathcal{S}^+$ used
in the conventional VC workflow. This choice isolated the effect of replacing
pre-set linearisations with real-time ones, and was adequate for the
preceding control tasks. Shot $54168$ deliberately departed from normal MAST-U
practice by assigning active feedback control to all seven RTVC shape
parameters simultaneously. This mode is normally avoided because several
shape parameters have strongly correlated PF coil responses. Asking the
controller to correct them independently can therefore make the local shape
matrix poorly conditioned, causing its pseudoinverse to convert shape
corrections into disproportionately large PF coil current requests. The
relevant limitation is the rank and conditioning of
$\mathcal{S}=\partial\bm{P}/\partial\bm{I}_{\text{act}}$.
This can be explicitly observed by comparing the quantity $\| \mathcal{V}_{\text{act}} \|_2$ in shot $54168$ with those in $53996$ (see \cref{fig:inverse_amplification}). Here $\| \mathcal{V}_{\text{act}} \|_2$ represents the norm of the VC matrix comprising the active control vectors.
Shot $53996$ demonstrates that norms of order $(1$--$2)\times 10^5$\,\si{\ampere\per\metre} are compatible with successful, stable shape control. 
Shot $54168$, however, produces significantly higher VC drives through much of the controlled period. Consequently, shape-control requests of a given magnitude are
translated into larger PF coil current requests, increasing sensitivity to
cross-coupling, measurement error, and model mismatch. This sustained
amplification is consistent with, and may have contributed to, the less smooth
shape tracking during shot $54168$. Although this comparison does not establish a
causal relationship, \cref{fig:inverse_amplification} shows that the amplification
is particularly pronounced close to the event at $\SI{0.50}{\second}$, when the
$\Rnose$ and $\Rs$ columns of the local shape matrix $\mathcal{S}$ become nearly
collinear. This near-degeneracy causes the pseudoinverse to amplify shape
corrections into large PF coil current requests. Damped or
Tikhonov-regularised inversion could limit this amplification, while
state-dependent conditioning could increase the damping when near-degenerate
shape directions are detected. RTVC already provides the evolving shape
matrix $\mathcal{S}$, so these changes affect only the small downstream
inversion and require neither emulator retraining nor changes to the
surrounding PCS architecture. They will be included in future RTVC
development.

Taken together, the results support a route towards simpler preparation of
plasma shape control. A single emulator trained over the relevant operating
space can provide local linearisations as a plasma discharge evolves, removing the
need to pre-compute bespoke phased VC schedules and carry out scenario-specific retraining of neural control policies. Routine use of RTVC will require a richer real-time
plasma state, a better-conditioned inverse, and validation over a wider
operational envelope. The central enabling step has nevertheless been
demonstrated: RTVCs can be integrated into an existing PCS and used to
control plasmas across substantially different scenarios without changing the
underlying control architecture.


\section*{Acknowledgements}

This work was funded by the Fusion Computing Lab collaboration (between UKAEA and STFC Hartree) and part funded by the EPSRC Energy Programme (EP/W006839/1).

For the purpose of open access, the authors have applied a Creative Commons Attribution (CC BY) licence to any author accepted manuscript version arising from this submission.
To obtain further information, please contact \texttt{publicationsmanager@ukaea.uk}.


\section*{Data availability}

The data supporting this work will be made available in accordance with the MAST-U data-access policy. 


\section*{Declarations}
The authors have no conflicts of interest to declare.



\begingroup
\small                        
\bibliographystyle{abbrvnat}  
\bibliography{references}  
\endgroup


\end{document}